\documentclass[twocolumn,aps,prl,amsmath,amssymb,superscriptaddress,showpacs]{revtex4-1}

\usepackage{graphicx}
\usepackage{graphics}
\usepackage{dcolumn}
\usepackage{bm}
\usepackage{color}
\usepackage{soul}
\usepackage{textcomp}
\usepackage{soul}
\usepackage{rotating}
\usepackage{setspace}
\usepackage[mathlines]{lineno}
\usepackage{textcomp}
\usepackage{mathcomp}

\newcommand{\affHav}{Department of Chemistry and Chemical Biology, Harvard University, Cambridge, MA 02138 USA.}

\usepackage{microtype}
\usepackage[breaklinks=true,hyperindex=true,
pdftitle={Quantum Origin of the 1/f-Noise in Chaotic Quantum Systems},
colorlinks=true,pagebackref=false,citecolor=blue,plainpages=false,pdfpagelabels,
linkcolor=blue,urlcolor=blue]{hyperref}

\begin{document}

\title{Origin of the $1/f^{\alpha}$-Spectral-Noise in Chaotic and Regular Quantum Systems}
\author{Leonardo A. Pach\'on} 
\thanks{Author to whom correspondence should be addressed.}
\affiliation{Grupo de F\'isica At\'omica y Molecular, Instituto de F\'{\i}sica,  
Facultad de Ciencias Exactas y Naturales, 
Universidad de Antioquia UdeA; Calle 70 No. 52-21, Medell\'in, Colombia.}
\author{Armando Rela\~no} 
\affiliation{Departamento de F\'isica Aplicada I and GISC, Universidad Complutense de Madrid, Avenida Complutense s/n, 28040 Madrid, Spain.}
\author{Borja Peropadre}
\affiliation{\affHav}
\affiliation{Quantum Information Processing group, Raytheon BBN Technologies, 10 Moulton Street, Cambridge, Massachusetts 02138, USA}
\author{Al\'an Aspuru-Guzik}
\affiliation{\affHav}

\begin{abstract}
Based on the connection between the spectral form factor and the probability to return, the 
origin of the $1/f^\alpha$-noise in fully chaotic and fully integrable systems is traced to the quantum 
interference between invariant manifolds of the classical dynamics and the dimensionality of those
invariant manifolds.
%
%
This connection and the order-to-chaos transition are analyzed in terms of the statistics of 
Floquet's quasienergies of a classically chaotic driving non-linear system.
An immediate prediction of the  connection established here is that in the presence of decoherence,
the spectral exponent $\alpha$ takes the same value, $\alpha=2$, for both, fully chaotic and fully 
integrable systems.

\end{abstract}

\date{\today}
\pacs{03.65.Sq, 31.15.Gy, 31.15.Kb}

\maketitle

{\it Introduction.}\textemdash
Quantum systems that in the classical limit are fully chaotic exhibit a variety of universal 
features  \cite{Haa10} such as level repulsion \cite{BGS84} or, in the semiclassical limit, a 
non-linear dependence on time of the spectral form factor $K(\tau)$ \cite{HO84}.
These seminal results were obtained on the basis of the predictions of Random Matrix Theory
(RMT) \cite{Meh91} and ramifications of the Gutzwiller trace formula  \cite{Gut67,Gut71}, 
respectively.
Only recently, a connection between these pioneering works was established in context of 
semiclassical periodic-orbit theory \cite{MH&04,HM&07}.

Based on a time-series perspective, a decade ago, it was discovered and proved that the spectral 
fluctuations of fully chaotic systems display $1/f$-noise whereas for fully integrable systems,
the spectral noise behaves as $1/f^2$ \cite{RM&02,GR&05,FG&04}.
By means of RMT, it is also possible to show that for KAM systems, mixed chaotic systems, 
chaos-assisted tunnelling \cite{DH81} makes the spectral noise to behave as $1/f^\alpha$ with
$1 < \alpha < 2$ \cite{Rel08}.
Therefore, the order-to-chaos transition is fully characterized by the spectral exponent $\alpha$
and contrary to the Dyson $\Delta_3(L)$ statistic, the exponent $\alpha$ quantifies the chaoticity 
of the system in a single parameter. 
Moreover, $\alpha$ is a natural measure of the fluctuation properties of a quantum system through 
the power spectrum.

However, having being developed on the basis of RMT, the $1/f^\alpha$ behaviour is the result
of statistical averages over the probability distribution of the elements of random matrices.
Therefore, it is not possible to interpret, e.g., the particular value of $\alpha$ for fully chaotic or 
integrable systems, in terms of invariant manifolds of the dynamics.
Since the average power noise that defines the $1/f^\alpha$ behaviour is a function of the spectral 
form factor $K(\tau)$ (see below), an interpretation is provided here on the basis of recent progress 
towards the identification of the classical invariant manifolds that contribute to the spectral form 
factor \cite{DP09}.
Specifically, by resolving the spectral form factor in phase space, it is shown that the particular 
value of $\alpha$ for fully chaotic and regular systems can be understood in terms of the 
dimensionality of the classical invariant manifold of the dynamics (one dimensional for isolated 
unstable periodic orbits and $N$-dimensional for regular tori) and their coherent quantum 
interference. 

The connection established here permits identifying the different values of the spectral
exponent $\alpha$ as a delicate interplay between quantum and classical signatures of 
the dynamics, namely, quantum interference and the dimensionality 
of classical invariant structures.
The consequences of this connection are manifold, e.g., it predicts that the
different value of the  spectral exponent for fully chaotic and fully integrable systems doest 
not survive in the classical limit.

{\it Spectral fluctuations: The average power noise and the 
spectral form factor.}\textemdash
The fluctuating parts of the level and accumulated level densities are denoted 
by $\tilde{\rho}(\epsilon)$ and $\tilde{n}(\epsilon)$, respectively.
Spectral fluctuations are analyzed in terms of the form factor $K(\tau)$
and the power spectrum $P^n(\tau)$, defined as the square modulus of the Fourier transform 
of $\tilde{\rho}(\epsilon)$ and $\tilde{n}(\epsilon)$, respectively.
For $\tau\neq0$, under the assumptions that 
$\langle\tilde{\rho}(\epsilon) \tilde{\rho}(\epsilon+\eta) \rangle\rightarrow 0$ faster than $1/\eta$ as $\eta\rightarrow \infty$
and for a large energy window $\Delta E\gg1$, it can be shown that \cite{FG&04}
\begin{equation}
\label{equ:PnKtau}
\frac{\langle | \hat{n}(\tau)|^2 \rangle}{\Delta E} = \langle P^n(\tau) \rangle  
= \frac{K(\tau)}{4\pi^2\tau^2},
\end{equation}
where $\langle \cdot \rangle $ stands for spectral averages whereas $ \hat{\cdot} $
does for Fourier transform of $ \tilde{\cdot}$.
The program developed in Refs.~\cite{RM&02,GR&05,FG&04} aims at introducing
a time series perspective to characterize the spectral noise of $\langle P^n(\tau) \rangle $.
The main idea behind this approach is to consider the sequence of energy levels as a 
discrete time series and study level correlations using tools from time-series analysis. 

{\it Time-series perspective of quantum chaos: The average power noise and the 
spectral form factor.}\textemdash
The analogy between the energy spectrum and a discrete time series is established in 
terms of the $\delta_q$ statistic \cite{RM&02}, defined as the deviation of the $(q+1)$-th 
level from its mean value. 
In terms of unfolded energy levels
$
\delta_q = \sum_{i=1}^q \left( s_i - \langle s \rangle \right) = \epsilon_{q+1} - \epsilon_1 - q,
$
where $s_i = \epsilon_{i+1} -\epsilon_i$, $\epsilon_i$ is the $i$-th unfolded level and
$ \langle s \rangle =1$ is the average value of $s_i$.

The unfolded energy levels are defined using the average accumulated level density 
$\bar{N}(E)$ as $\epsilon_i = \bar{N}(E_i)$. 
This mapping is needed to remove the main trend defined by the smooth part of the 
level density and compare between the statistical properties of the spectral fluctuations 
of different systems or different parts of the same spectrum. 
In the language of time series analysis, the unfolding mapping is a procedure for making 
stationary the discrete time series defined by the $\delta_q$, its average and fluctuations 
not depending on time.
Sampling $\tilde{n}(\epsilon)$ for integer values of the energy leads to the discrete function
$\tilde{n}_q(\epsilon)$ with averaged power spectrum $\langle P_k^n\rangle$ and Fourier 
transform is given by $\hat{n}_k = D_\mathcal{H}^{-\frac{1}{2}} \sum_{q=-\infty}^{\infty} 
\hat{n}\left( k/D_{\mathcal{H}} +q \right)$, with $k=1,2,\ldots,D_{\mathcal{H}}-1$.
$D_{\cal{H}}= \Delta E/\langle d\rangle$ is the effective dimension of the Hilbert space 
${\cal{H}}$ and $\langle d\rangle$ denotes the mean spectral density for a finite range $\Delta E$.
The averaged power noise of $\delta_k$ is related to $\langle P_k^n\rangle$ by
$\langle P_k^{\delta} \rangle = \langle P_k^n \rangle -\frac{1}{12}$ for chaotic systems and 
by $\langle P_k^{\delta} \rangle = \langle P_k^n \rangle$ for regular systems.
If $D_\mathcal{H} \gg 1$ and $k\ll D_\mathcal{H}$, 
$\langle P_k^{\delta} \rangle_\beta = D_\mathcal{H}/(2\beta\pi^2k)$ for chaotic systems belonging 
to the three $\beta=\{1,2,4\}$ classical RMT  (for fully chaotic 
systems)  whereas $\langle P_k^{\delta} \rangle = D_\mathcal{H}^2 /(4\pi^2 k^2)$ 
for integrable systems.
Thus, for small frequencies, the excitation energy fluctuations exhibit $1/f$ ($\sim 1/k$) noise in chaotic 
systems and $1/f^2$ ($\sim 1/k^2$) noise in integrable systems \cite{FG&04}.

{\it Interference of time-domain scars: Spectral form factor and probability 
to return}\textemdash
The key quantity that allows for the identification of  the contribution of classical invariant manifolds 
to $K(\tau)$ is the probability to return $P^{\mathrm{qm}}_{\mathrm{ret}}(t)$ \cite{DS91,Dit96,DP09}.
To make a clear connection with the classical invariant manifolds of the underlying 
classical dynamics, it is convenient to express the return probability in terms of 
phase-space objects.
To do so, introduce the Weyl representation of quantum mechanics
\cite{Wey50}, which assigns a phase space function $O(\mathbf{p},\mathbf{q})$ to an 
operator $\hat{O}$.
For the density operator $\hat{\rho}(t)$ at time $t$, the Weyl transform defines the Wigner 
function
$\rho_{\mathrm{W}}(\mathbf{r},t) 
= \int \mathrm{d}^N q' \exp(-\mathrm{i} \mathbf{p}\cdot \mathbf{q}'/\hbar)
\langle \mathbf{q} +  \mathbf{q}'/2| \hat{\rho}(t)| \mathbf{q}- \mathbf{q}'/2\rangle$
with $\mathbf{r} = (\mathbf{p},\mathbf{q})$ a vector in $2N$-dimensional
phase-space.
The propagator $G_{\mathrm{W}}(\mathbf{r}'',t'';\mathbf{r}',t')$ of the Wigner function 
evolves the Wigner function from $t'$ to $t''$, 
$\rho_{\mathrm{W}}(\mathbf{r}'',t'')  = 
\int \mathrm{d}^{2N}r G_{\mathrm{W}}(\mathbf{r}'',t'';\mathbf{r}',t')
\rho_{\mathrm{W}}(\mathbf{r}',t') $ and has a clear classical analog, namely, the 
Liouville propagator \cite{Pri62}.

The quantum probability to return can be expressed as a trace over phase space 
of the propagator of the Wigner function, namely,
$P^{\mathrm{qm}}_{\mathrm{ret}}(t) = \int \mathrm{d}^{2f}r_0 
G_{\mathrm{W}}(\mathbf{r}_0,t;\mathbf{r}_0)$, with $t=t''-t'$. 
For $t\gtrsim t_{\rm H}/D_{\cal{H}}$, the form factor is related to the quantum return probability 
by \cite{DP09}
\begin{equation}
\label{equ:FrmFctrRtrnPrblty}
D_{\mathcal{H}} K(\tau) = \int \mathrm{d}^{2f}r G_{\mathrm{W}}(\mathbf{r},t;\mathbf{r},0) 
= P^{\mathrm{qm}}_{\mathrm{ret}}(t),
\end{equation}
with $\tau=t/t_{\mathrm{H}}$ and $t_{\rm H} = h\langle d\rangle$ denotes the Heisenberg time.
Remarkably, before tracing, the quasiprobability density to return $G_{\mathrm{W}}(\mathbf{r},t;\mathbf{r},0)$
allows for the identification of the manifolds that contributed to the form factor (see Fig.~\ref{fig:diagWgnrprpgtr}
below).
At the semiclassical level, besides the classical invariant manifolds 
with period $T^{\mathrm{p}}$ and invariant manifolds with period $T=T^{\mathrm{p}}/l$, 
being $l$ and integer, also sets of midpoints between them contribute \cite{DP09}. 
These midpoint manifolds constitute important exceptions from a continuous convergence 
in the classical limit of the Wigner towards the Liouville propagator \cite{DP09} and, as shown 
below, are responsible for the different functional form of the spectral noise in chaotic and regular 
systems.

\begin{figure*}
\begin{tabular}{cc}
\includegraphics[width=0.755\columnwidth]{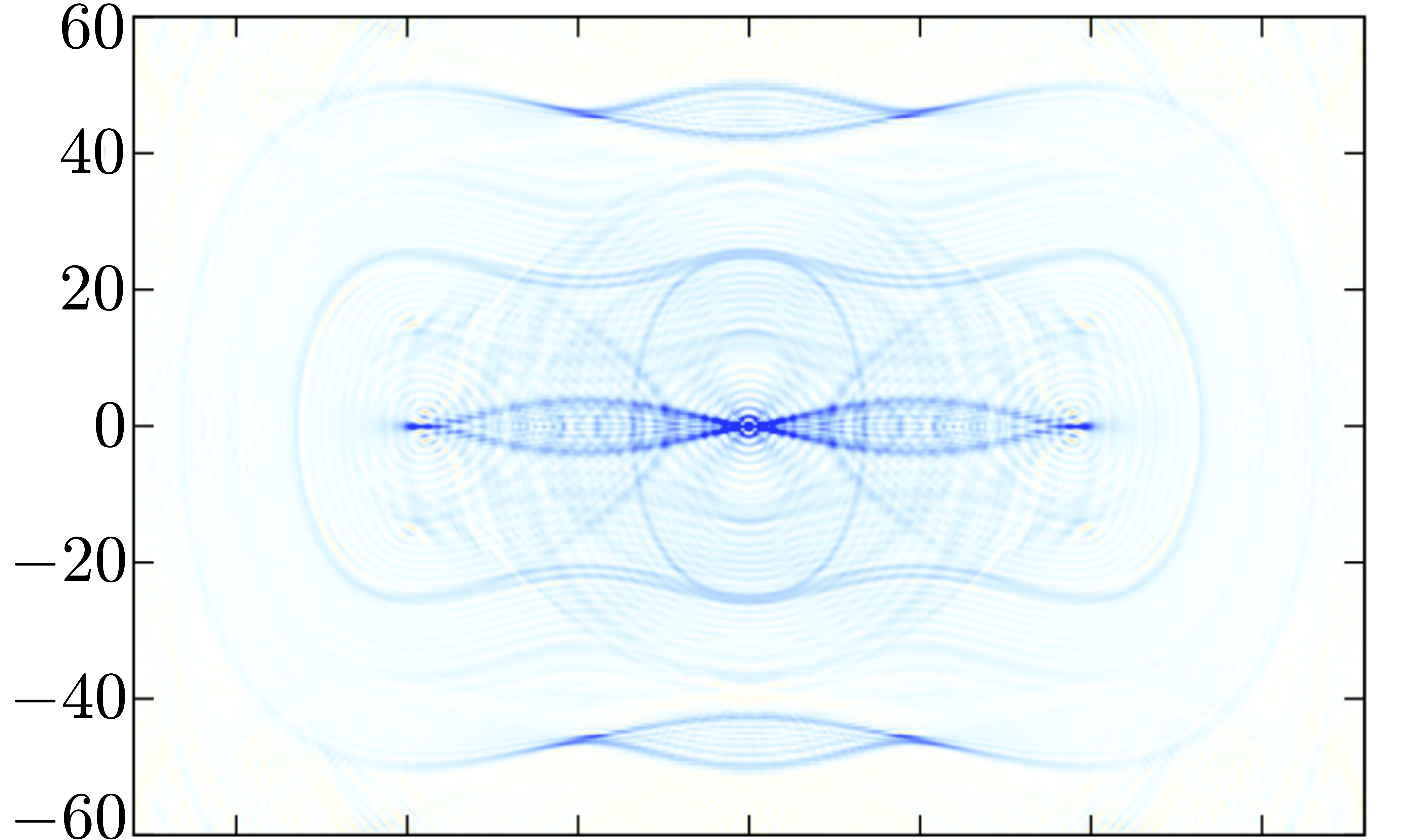}
&
\includegraphics[width=0.6975\columnwidth]{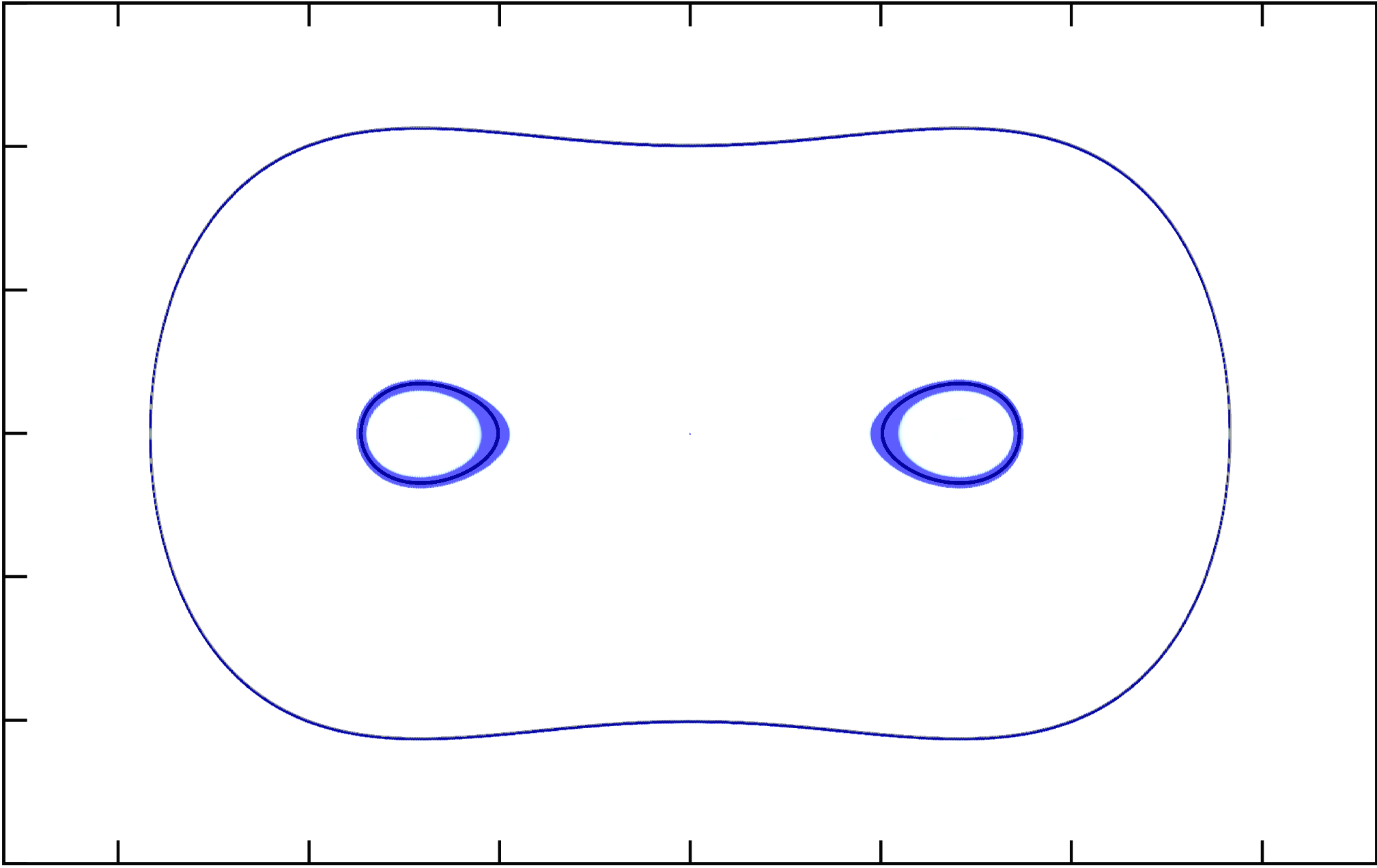}
\\
\includegraphics[width=0.75\columnwidth]{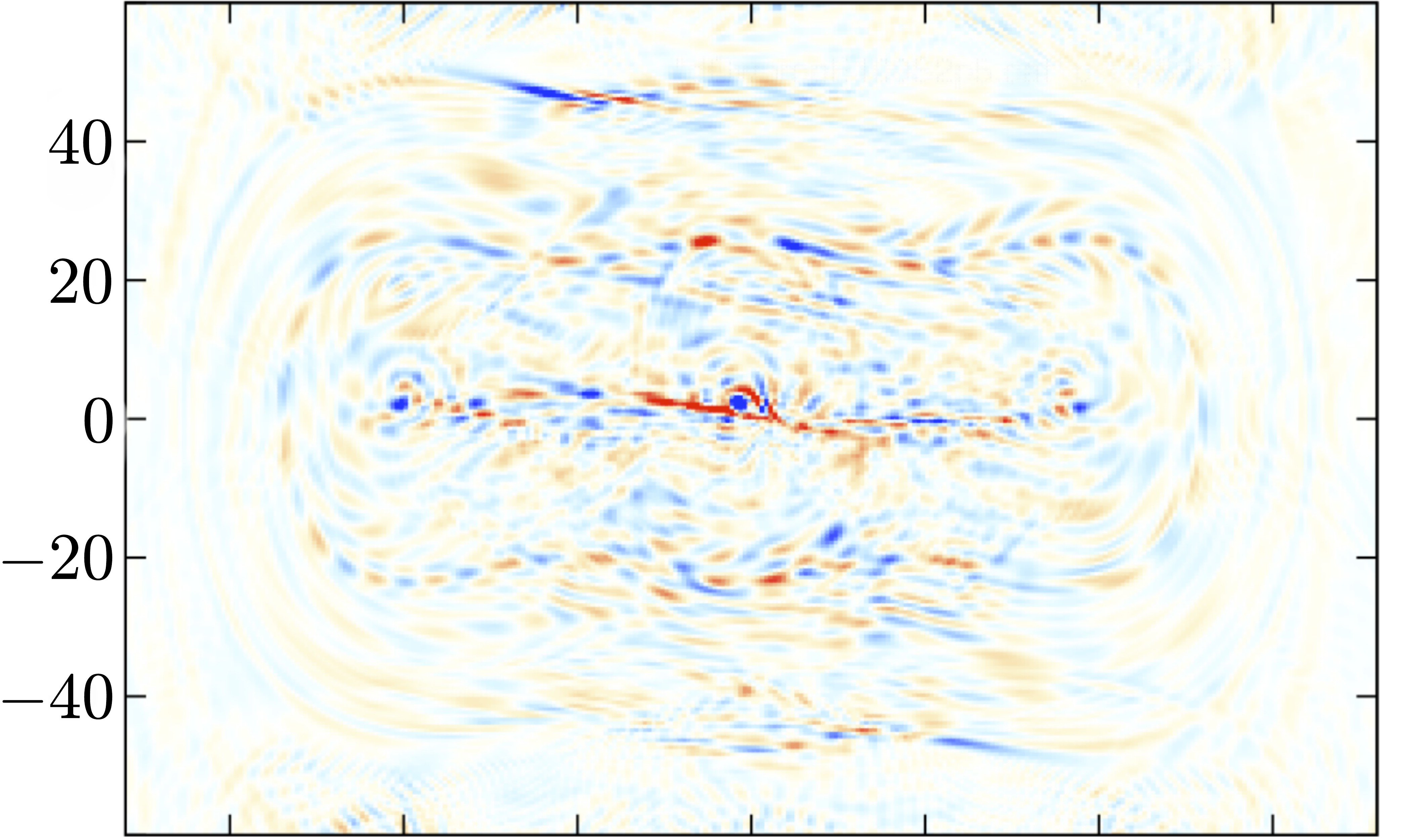}
&
\includegraphics[width=0.7\columnwidth]{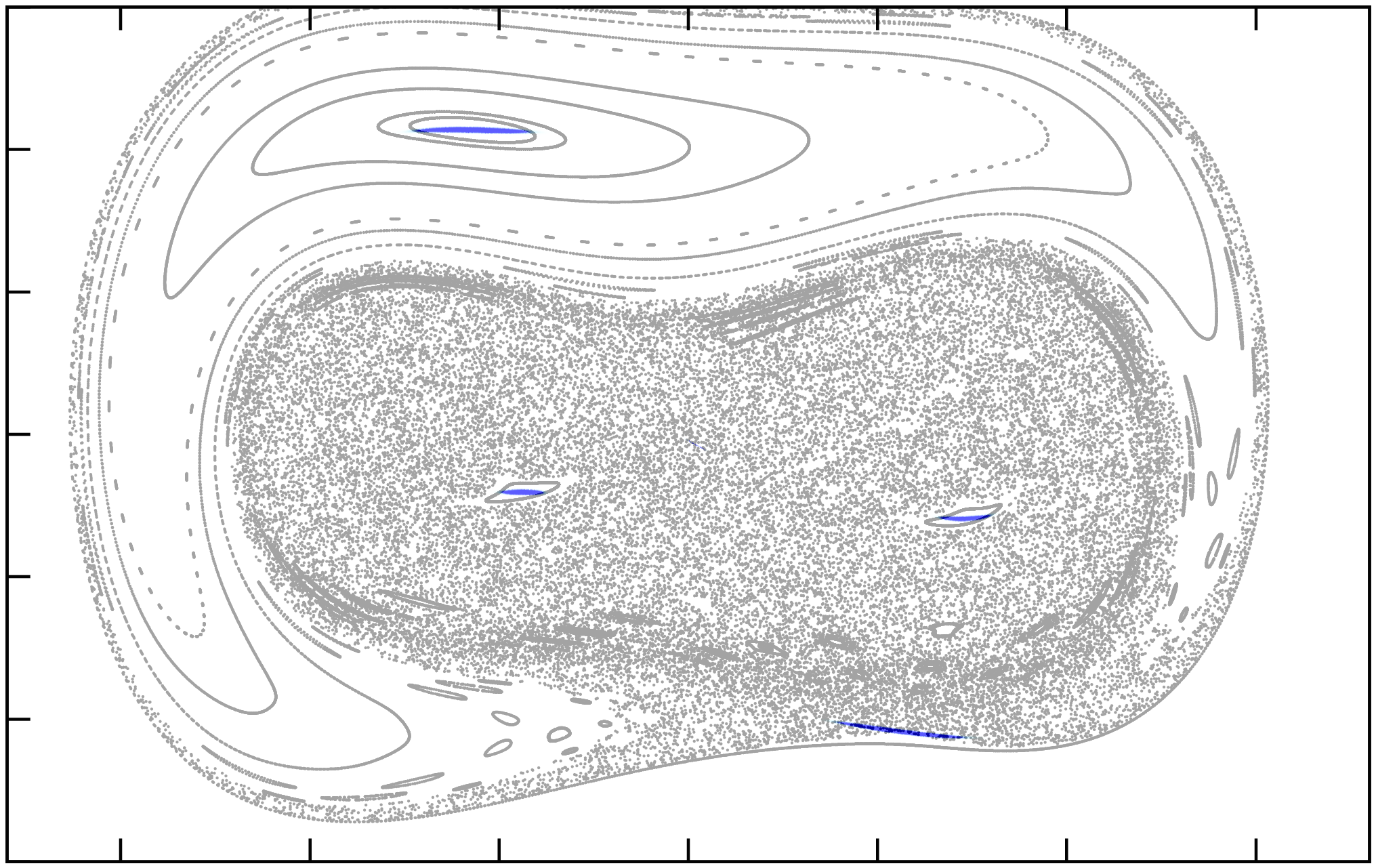}
\\
\includegraphics[width=0.8\columnwidth]{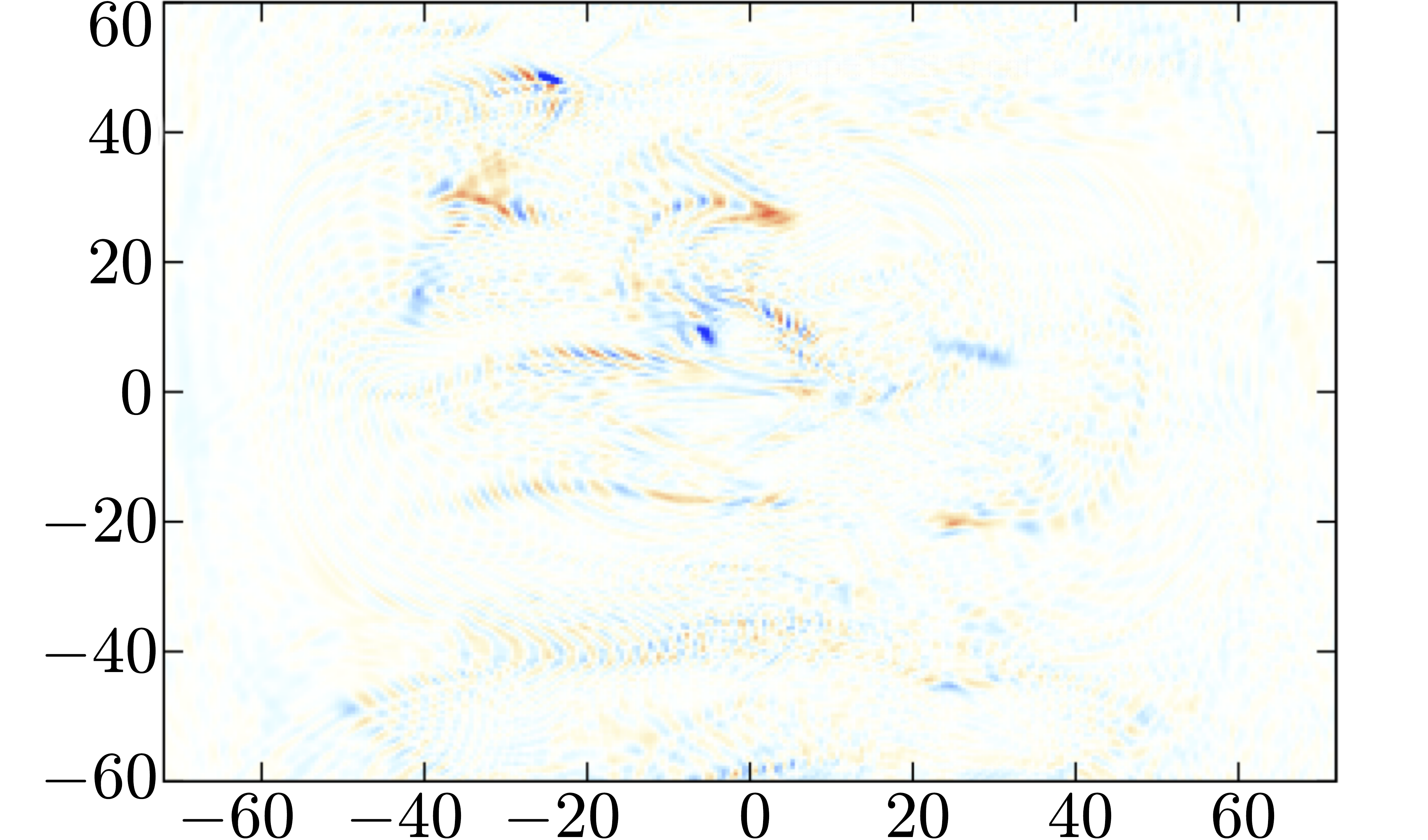}
&
\includegraphics[width=0.7\columnwidth]{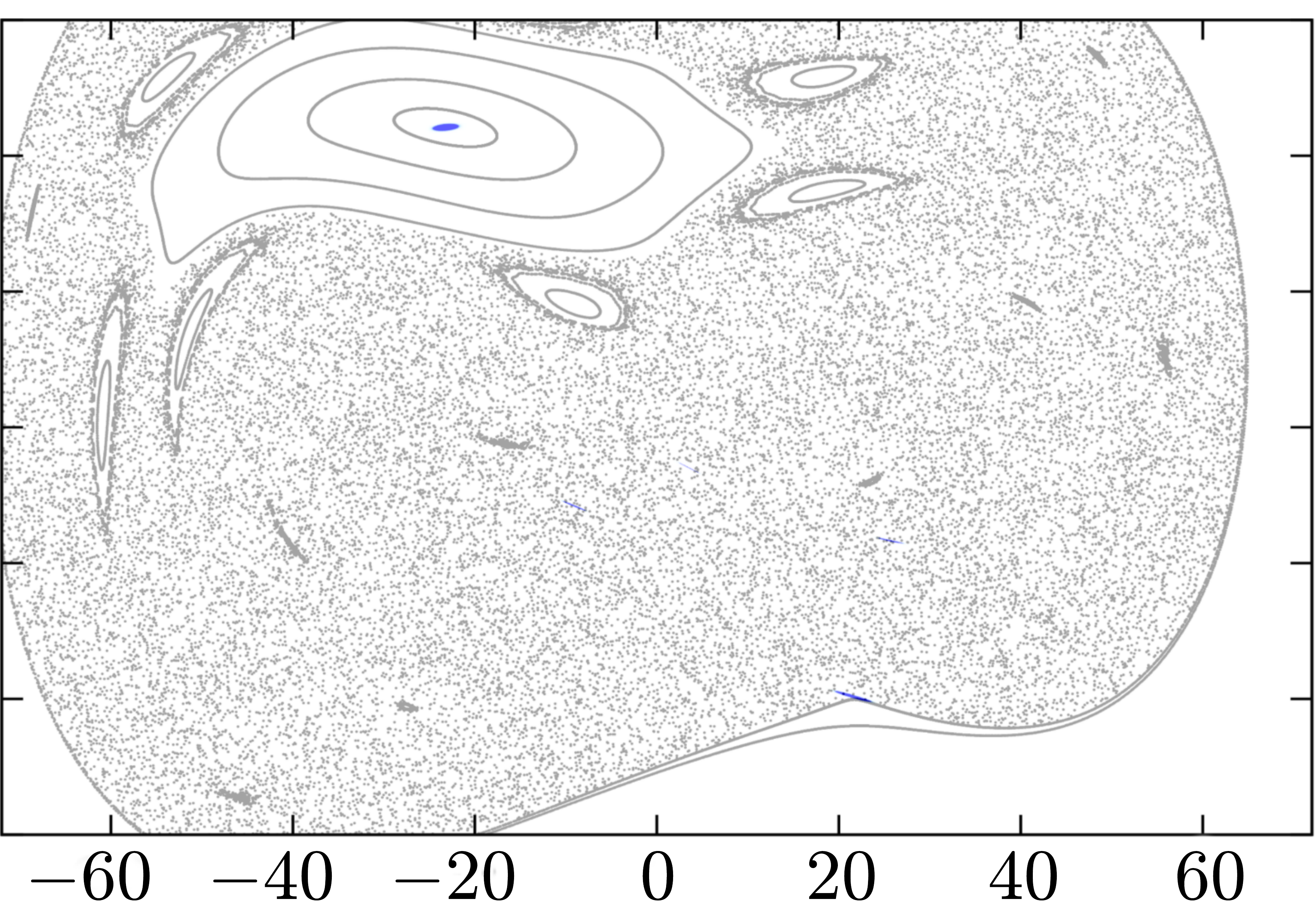}
\end{tabular}
\caption{Quantum (l.h.s. panel) and classical (r.h.s. panel) diagonal propagators $G({\bf r},t;{\bf
r},0)$ for the harmonically driven quartic oscillator at $t = T \equiv
2\pi/\Omega$, with $\omega_0 = 1.0$, $\Omega = 0.95$,
$E_{\rm b} = 100.0$, $\phi=\pi/3$ with $S = 0$ (upper panels), $S = 2.5$ (central panels) and
$S = 10$ (lower panels).
For better understanding of the midpoint contributions in the diagonal propagator, Poincare surface 
of sections are shown behind of the classical diagonal propagator.
}
\label{fig:diagWgnrprpgtr}
\end{figure*}

{\it Probability to return and the average power noise}\textemdash
The connection between the averaged power-spectrum of the spectral fluctuations and the 
invariant manifolds of the classical dynamics, and their quantum interferences, is established 
from the comparison between 
Eqs.~(\ref{equ:PnKtau}) and (\ref{equ:FrmFctrRtrnPrblty})
\begin{equation}
\label{equ:PnRtnPrb}
 \langle P^n(\tau) \rangle  = D_{\mathcal{H}}^{-1} (2\pi \tau)^{-2}  
 P^{\mathrm{qm}}_{\mathrm{ret}}(t)
\end{equation}
Because this identity does not rely on any semiclassical approximation, it is 
exact and holds for finite- and infinite-dimensional Hilbert spaces.
Moreover, because it is formulated at the level of the statistical operator $\hat{\rho}$ 
and not at the level of elements of the projective Hilbert space, it holds for unitary 
as well as non-unitary dynamics.

As stated above, the calculation of $\langle P_k^{\delta} \rangle$ requieres the unfolding 
of the energy leves.
Here, that unfolding needs to be reinterpreted and calculated at the level of the return 
probability, which is defined as a direct trace over phase space of the diagonal propagator 
$G_{\mathrm{W}}(\mathbf{r},t;\mathbf{r},0)$.
Thus, the subtraction of the main trend translates here into to the subtraction 
of the classical contribution, i.e., it is assumed that the quantum propagator 
can be accounted for by the superposition of the classical propagator plus quantum fluctuations 
\cite{DVS06,DP09,DGP10,PID10}.
Thus, define the quantities
$\Delta P_{\mathrm{ret}}(t) = P^{\mathrm{qm}}_{\mathrm{ret}}(t) - P^{\mathrm{cl}}_{\mathrm{ret}}(t)$
and 
$\langle \Delta P^n(\tau) \rangle = D_{\mathcal{H}}^{-1} (2\pi \tau)^{-2} 
\Delta P_{\mathrm{ret}}(t)$.
In the semiclassical limit,
\begin{equation}
\label{equ:SmclssRtnPrb}
P^{\mathrm{qm}}_{\mathrm{ret}}(t)  \approx
 \begin{cases}
D_{\mathcal{H}} (2/\beta) \tau P^{\mathrm{cl}}_{\mathrm{ret}}(t), & \mathrm{for\,fully\,chaotic\, systems},
\\
D_{\mathcal{H}}  P^{\mathrm{cl}}_{\mathrm{ret}}(t), & \mathrm{for\,integrable\, systems},
 \end{cases}
\end{equation}
where no degeneracies are considered for the integrable case \cite{Dit96}.
So that, for $D_{\mathcal{H}}\gg 1$,
\begin{equation}
\label{equ:SmclssPnRtnPrb}
\frac{\langle \Delta P^n(\tau) \rangle}{P^{\mathrm{cl}}_{\mathrm{ret}}(t)}  \approx
 \begin{cases}
(2\pi^2 \beta)^{-1} \tau^{-1}, & \mathrm{for\,fully\,chaotic\, systems},
\\
(4\pi^2 )^{-1} \tau^{-2}, & \mathrm{for\,integrable\, systems}.
 \end{cases}
\end{equation}
$\langle \Delta P^n(\tau) \rangle/P^{\mathrm{cl}}_{\mathrm{ret}}(t)$ measures deviations 
from the main trend, classical contributions, normalized by the classical return probability.
From Eq.~(\ref{equ:SmclssPnRtnPrb}), it is clear that the description in terms of the return 
probability provides consistent results with the time-series perspective developed in 
Refs.~\cite{RM&02,GR&05,FG&04}, i.e.,  deviations of the averaged power spectrum from 
the main trend behave as $1/\tau^\alpha$ with $\alpha=1$ for chaotic and $\alpha=2$ for 
integrable systems, respectively.
The results formally coincides after, as defined above, replacing $\tau$ by  
$k/D_{\mathcal{H}}$ .

The main advantage of the present formulation relies on the possibility of interpreting 
the origin of the different values of the exponent $1 \leq \alpha\leq 2$.
As shown above, the different nature of the spectral noise relies on the particular functional 
dependence of the quantum return probability on $\tau$ [see Eq.~(\ref{equ:SmclssRtnPrb})]. 
Therefore, this particular dependence relies on the different nature of classical invariant manifolds 
that contribute to the quantum return probability \cite{DP09}.
Specifically, it is understood in terms of midpoints manifolds showing up from the interference 
of periodic invariants of the dynamics (see, e.g., Fig.~\ref{fig:diagWgnrprpgtr} and description 
below).
For regular systems, the number and size of these manifolds scale with time the
same way as that of the underlying tori \cite{HO84,DP09}, so that no $\tau$ factor arises between 
quantum and classical return probabilities in Eq.~(\ref{equ:SmclssRtnPrb}).
This situation in turn reflects the fact that periodic tori form $N$-dimensional surfaces in phase 
space and are space filling, e.g., in position space.
In contrast, isolated periodic orbits remain one-dimensional subsets independently of the number 
of freedoms, this dimensionality property is in turn responsible for the emergence of the
$\tau$ factor for fully chaotic systems \cite{DP09}. 
There is therefore qualitatively ``more room'' available for midpoint manifolds
in the latter case than in the former.
%

{\it Example}\textemdash
Traditionally, the study of spectral fluctuations have been performed in nuclear systems
or 2D billiards \cite{RM&02,GR&05,FG&04}.
However, because the dimension of the Wigner propagator is four times the real space 
dimension, a phase-space characterization of the invariant manifolds in this object is not feasible 
for those systems. 
For this reason, a one-dimensional driven mixed chaotic systems, prototypical in, e.g., coherent 
destruction of tunnelling \cite{GD&91}, is considered here, namely,
$H = p^2/2m - m \omega^2 q^2/4 + m^2 \omega^4q^4/64 E_\mathrm{b}
+S \cos (\Omega t + \phi)$.
$E_{\mathrm b}$ denotes, roughly, the number of tunnelling doublets and $S$ stands for the strength 
of the driving force.

Because of the periodicity of the driving force, spectral fluctuations are analyzed for Floquet's 
quasienergies, that are eigenvalues of the unitary-time evolution operator $\hat{U}(t)$ over
one period of driving $T=2\pi/\Omega$.
To the best of our knowledge, this is the first time that a characterization of the spectral noise
is performed for a driven mixed chaotic systems and thus, some comments are in order. 
The unitarity of the time-evolution operator implies that its eigenvalues are of unit magnitude 
and therefore, they can be conveniently written as $\exp(\mathrm{i} E_\alpha T/\hbar)$, being 
$E_\alpha$ Floquet's quaisenergies. 
$E_\alpha$ is defined modulo integer multiples of $\hbar \Omega$, namely, 
$E_\alpha = E_{n,l} = E_{n,0} + l \hbar \Omega$, $n=0,1,2, ...$ and 
$l=0,\pm 1, \pm 2, ...$
\begin{figure}[h]
\includegraphics[width=\columnwidth]{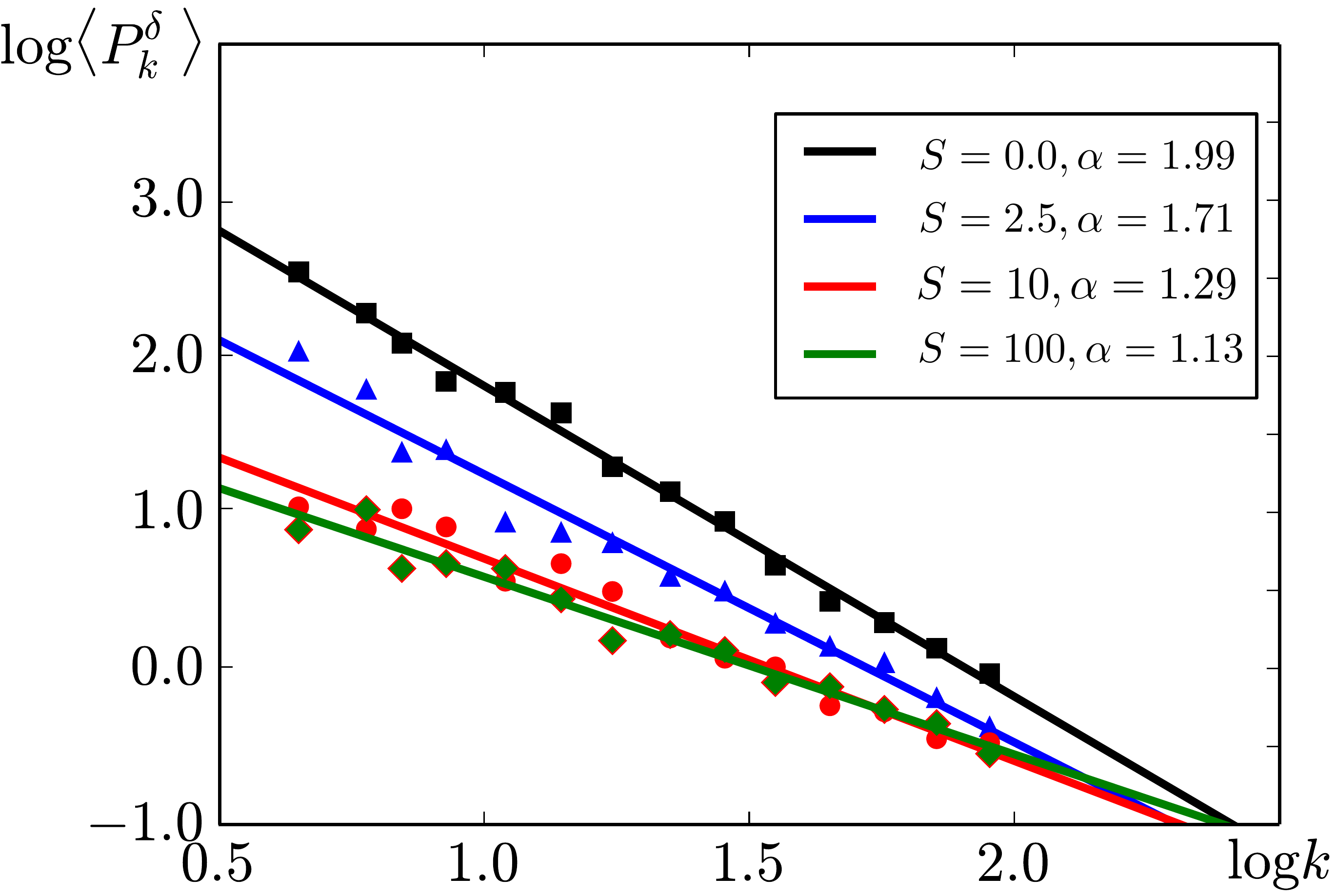}
\caption{Theoretical power spectrum of the $\delta_q$ function. 
Parameter values are as in Fig.~\ref{fig:diagWgnrprpgtr}. }
\label{fig:spctrlflcttns}
\end{figure}
For the spectral statistics, only the quasienergies in the first Brillouin's zone, $l=0$, are 
considered, so that $-\hbar \Omega < E_n <  \hbar \Omega$.

Before discussing the spectral features of this system, it is instrumental to have a qualitatively 
idea about the underlying manifolds that will determined the spectral exponent $\alpha$. 
Figure~\ref{fig:diagWgnrprpgtr} depicts the quantum quasi-probability density $G_{\mathrm{W}}(\mathbf{r},t,\mathbf{r},0)$
for the driven doble well potential considered above for zero driving ($S=0$, upper panel), strong driving  
($S=2.5$, central panel) and ultra strong driving ($S=10$, lower panel).
For the classical dynamics of the undriven case, there exists three periodic orbits of period $T$ 
that can be clearly seen in the diagonal classical propagator (l.h.s. of Fig.~\ref{fig:diagWgnrprpgtr}).
Thee is also a family of orbits whose period is a rational fraction of $T$, e.g., $T/2$, that is located 
outside of the domain of the plot.
The interference of these manifold is clearly visible in the upper panel of Fig.~\ref{fig:diagWgnrprpgtr}.
In the presence of driving, these continuous manifold are replaced by a set of unstable elliptical and 
hyperbolic periodic points (see central and lower panels in in Fig.~\ref{fig:diagWgnrprpgtr}). 
Remarkably, the quantum interference between these manifold (contributions from midpoints between
classical invariant manifolds) is also clearly visible in the central and lower panels of 
Fig.~\ref{fig:diagWgnrprpgtr}.

Figure \ref{fig:spctrlflcttns} depicts the functional dependence of $\langle P^{\delta}_k \rangle$ on $k$ for 
Floquet's quasienergies for $S=0$ ($\alpha=1.99$), $S=2.5$ ($\alpha=1.71$), $S=10$ 
($\alpha=1.29$) and $S=100$ ($\alpha=1.13$). 
Despite the KAM nature of the system at hand, the spectral fluctuations exhibit a clear
$1/f^\alpha$ dependence.
This feature of Floquet's quasienergies supports the evidence found in the Robnik billiard 
\cite{GR&05}, and are in sharp contrast to the conventional expectation that in the strict 
semiclassical limit spectral fluctuations of mixed chaotic systems cannot follow a power law
\cite{GR&05,Rel08}.

{\it Discussion}\textemdash
In this Letter, by establishing a connection between the power noise and the probability to 
return [see Eqs.~(\ref{equ:PnRtnPrb}, \ref{equ:SmclssPnRtnPrb})], the origin of the $1/f^\alpha$ 
noise in quantum systems (for $\alpha=1,2$) was track to the interference and dimensionality 
of classical invariant manifolds of the regular and chaotic dynamics.
In the process, the main trend of the power spectrum was associated to the classical 
contribution to the quantum dynamics, so that $\langle \Delta P^n(\tau) \rangle/P^{\mathrm{cl}}_{\mathrm{ret}}(t)$
measures purely quantum fluctuations. 

The extreme values of the $\alpha$ parameter, $\alpha=1$ and $\alpha=2$, are the result of the 
$\tau$ dependence of the quantum return probability [see, Eq.~(\ref{equ:SmclssRtnPrb})].  
Because the connections stablished above are valid in general, it suggest that the fractional 
behaviour of the spectral noise, $1<\alpha<2$ emerges for the interference between regular 
and chaotic invariant manifolds, however, an analytic account of this fact remains as a challenge.

This same connection allows for the immediate prediction that in the presence of decoherence, 
the spectral coefficient $\alpha$ takes the same value,  $\alpha=2$, for classically chaotic and 
classically regular systems.
This follows from the fact that in the presence of decoherence, the quantum return probability
behaves equally for both, integrable and chaotic systems \cite{Bra99}.
In a nutshell, decoherence removes the interference between invariant manifolds so that
the additional coherent contributions to the form factor discovered in Ref.~\cite{DP09} are
not present anymore.
Work along this line will be reported soon \cite{Pac17}.

The approach presented here can be extended to uncover the invariant 
manifolds responsible for the behaviour of the power spectrum of energy level fluctuations
very recently discussed in the nonperturbative analysis of the  in fully 
chaotic quantum structures was reported \cite{ROK17}.

\begin{acknowledgements}
{\it Acknowledgements}\textemdash
Discussions with Thomas Dittrich are acknowledged with pleasure.
This work was supported by the the \emph{Comit\'e para el Desarrollo de la Investigaci\'on} 
-CODI-- of Universidad de Antioquia, Colombia under the Estrategia de Sostenibilidad 
2016-2017 and by \textit{Colombian Institute for the Science and Technology Development} 
--COLCIENCIAS-- under grant number 111556934912.
AR is supported by Spanish Grants No. FIS2012-35316 and No. FIS2015-63770-P (MINECO/FEDER).
\end{acknowledgements}

\bibliography{pqnprl}

\end{document}